\documentclass[twocolumn,floatfix,prd,authoryear,nofootinbib,aps,10pt,tightlines,eqsecnum]{revtex4-2}
\usepackage{bm}
\usepackage{newtxmath,newtxtext}
\usepackage{graphicx}
\usepackage{textcomp}
\usepackage{float}
\usepackage{booktabs}
\usepackage{dcolumn,enumerate}
\usepackage{ragged2e}
\usepackage{hyperref}

\hypersetup{colorlinks=true, linkcolor=blue, urlcolor=blue, citecolor=blue}

\usepackage{xcolor}
\usepackage{epsfig}
\usepackage{caption}
\usepackage{appendix}
\usepackage{subcaption}
\usepackage{graphicx} 
\usepackage{orcidlink}
\begin{document}

\title{Probing Kerr black hole in a uniform Bertotti-Robinson magnetic field through astrophysical quasi-periodic oscillations}

\author{Hamza Rehman $^{a, b, c}$\orcidlink{0009-0004-1553-9022}}
\email{hamzarehman244@zjut.edu.cn}
\author{Sanjar Shaymatov $^{d, e, f}$\orcidlink{0000-0002-5229-7657}}
\email{sanjar@astrin.uz}
\author{Saddam Hussain ${}^{a, b}$\orcidlink{0000-0001-6173-6140}}
\email{saddamh@zjut.edu.cn}
\author{Tao Zhu $^{a, b}$\orcidlink{0000-0003-2286-9009}}
\email{Corresponding author: zhut05@zjut.edu.cn}

\affiliation{${}^{a}$ Institute for Theoretical Physics and Cosmology, Zhejiang University of Technology, Hangzhou 310023, China}
\affiliation{${}^{b}$ United Center for Gravitational Wave Physics (UCGWP), Zhejiang University of Technology, Hangzhou, 310023, China}
\affiliation{${}^{c}$ Center for Theoretical Physics, Khazar University, 41 Mehseti Str., Baku, AZ1096, Azerbaijan}
\affiliation{${}^{d}$ Institute of Fundamental and Applied Research, National Research University TIIAME, Kori Niyoziy 39, Tashkent 100000, Uzbekistan}
\affiliation{${}^{e}$ Tashkent University of Applied Sciences, Gavhar Str. 1, Tashkent 100149, Uzbekistan}
\affiliation{${}^{f}$ Tashkent State Technical University, 100095 Tashkent, Uzbekistan}

\date{\today}

\begin{abstract}

In this study, the behavior of high-frequency quasi-periodic oscillations (QPOs) is investigated around a Kerr black hole immersed in a uniform Bertotti-Robinson magnetic field. The motion of the test particle is analyzed by determining the geodesic equations and evaluating the corresponding orbital, radial, and vertical epicyclic frequencies. These fundamental frequencies are used to construct the theoretical framework of QPO models based on parametric and forced resonance mechanisms. Observational data obtained from several black hole X-ray binaries (GRO J1655--40, XTE J1550--564, XTE J1859+226, GRS 1915+105, H1743--322, M82~X--1, and Sgr~A$^{*}$) are used to constrain the black hole parameters through Bayesian inference and Markov chain Monte Carlo analyses. For the X-ray binaries GRO J1655-40, GRS 1915+105, H1743-322, and M82~X-1, nonzero values of the dimensionless parameter $b=Bm$ are obtained at the $68\%$ confidence level within the framework of the parametric resonance model, while only upper bounds at the $90\%$ confidence level are obtained for the remaining sources. In contrast, in the case of the forced resonance model, only an upper bound at the $90\%$ confidence interval is obtained for the magnetic field parameter for all considered X-ray binary sources. The analysis indicates that the value of the magnetic field parameter is small but not negligible, producing minor modifications to particle dynamics and epicyclic frequencies. The influence of the magnetic field is further examined through the properties of the innermost stable circular orbit and the radiative properties of the thin accretion disk, including the energy flux and temperature profiles, within the allowed parameter range inferred from the Markov chain Monte Carlo analysis. These results demonstrate that the Kerr–Bertotti–Robinson spacetime can produce observable deviations from the Kerr black hole in X-ray observations. In particular, QPOs can provide useful constraints on the magnetic field parameter and the physical properties of the considered spacetime.

\end{abstract}

\maketitle

\section{Introduction}

The study of black hole (BH) dynamics is a fundamental aspect of contemporary theoretical and astrophysical investigations, providing a deep understanding of the interplay among gravity, quantum mechanics, and astrophysical phenomena. A significant advance in gravitational physics was made when the LIGO collaboration detected gravitational waves coming from binary BH mergers, achieving a groundbreaking validation of general relativity  \cite{LIGOScientific:2016aoc}. This accomplishment was followed by landmark observations by the Event Horizon Telescope (EHT) that captured the first images of supermassive BHs, first in the galaxy M87* and later in the core of the Milky Way, Sagittarius A* (SgrA*) \cite{EventHorizonTelescope:2019dse, EventHorizonTelescope:2019uob, EventHorizonTelescope:2019jan, EventHorizonTelescope:2019ths, EventHorizonTelescope:2019ggy, EventHorizonTelescope:2019pgp}. Apart from this direct imaging technique, quasi-periodic oscillations (QPOs) found in X-ray binary systems serve as an important tool to analyze the behavior of particles in the vicinity of spacetime geometry.

The discovery of QPOs as a distinctive astrophysical phenomenon in the 1980s \cite{Samimi:1979si} provided a remarkable opportunity to probe gravity in the strong field and to gain deeper insight into BH spacetime geometry through high-precision X-ray timing observations of BH X-ray binaries \cite{Stella:1998mq, Stella:1997tc}. A typical X-ray binary system consists of a donor star and a compact object, such as a BH or a neutron star. The compact object gravitationally accretes material from its companion, forming an accretion disk that becomes heated through viscous dissipation and consequently emits intense X-ray radiation. This enables us to study the relativistic motion of matter in intense gravitational fields and to investigate the highly compact matter that constitutes neutron stars. Observations of QPOs in the X-ray emission from celestial objects allow us to monitor the accretion flow onto the compact object through rapid temporal variations. The images of these frameworks provide an average angular resolution at the sub-nanoarcsecond level, exceeding that of current imaging instruments \cite{Ingram:2019mna, Remillard:2006fc}.

The QPOs are explained by several theoretical models, such as the relativistic precession model (RP), the warped disk (WD) model, parametric resonance (PR) models, and resonance models \cite{Stella:1997tc, Stella:1998mq, 1999ApJ...524L..63S, Cadez:2008iv, Kostic:2009hp, Germana:2009ce, Kluzniak:2002bb, Abramowicz:2003xy, Rebusco:2004ba, Nowak:1996hg, Torok:2010rk, Torok:2011qy, Xamidov25PDU....4701805X,Shaymatov23ApJ,Kotrlova:2020pqy,Xamidov25EPJC...85.1193X}. Such models link observable temporal aspects with the small oscillations of particles in a strong gravitational field, thereby enabling QPOs to probe the relativistic motion of particles near the compact object. This study focuses on the PR model, which involves nonlinear and parametric resonances between verticle and radial modes, while the FR model, in which resonances arise due to internal or external perturbations in the disk. Using these models, we will investigate how matter accretion triggers the QPO frequency in X-ray binaries. 

QPOs are observed in X-ray binaries around compact objects, providing valuable information about relativistic effects in strong gravitational fields. The QPO models were initially developed to describe the QPOs observed in neutron-star X-ray binaries but later extended to stellar-mass and supermassive BHs \cite{PhysRevLett.82.17}. Since a BH possesses a strong gravitational field, it serves as an ideal laboratory for investigating the spacetime geometry around the BH \cite{Motta:2013wga}.
QPOs models have been used to test the no-hair theorem and to study deviations from Kerr geometry (e.g., GRO J1655-40 and other BH candidates, modified-gravity scenarios, wormholes, and nonlinear electrodynamics) \cite{Allahyari:2021bsq, Banerjee:2022chn, Bambi:2012pa, Bambi:2013fea, Deligianni:2021ecz, Deligianni:2021hwt, Maselli:2014fca, Wang:2021gtd, Uktamov25EPJC...85.1432U,Jiang:2021ajk, Ashraf:2025lxs, Yang:2025aro, Guo:2025zca, Yang:2024mro, Liu:2023ggz, DeFalco:2023kqy, Bambi:2022dtw,Xamidov26JCAP...01..044X, Liu:2023vfh}. Furthermore, the test particles' motion and the resulting epicyclic frequencies around BHs have been analyzed in \cite{Shaymatov:2020yte,Dasgupta:2025fuh, Rayimbaev-Shaymatov21a,Banerjee:2021aln, Alloqulov25JHEAp,Jumaniyozov:2025wcs, Shaymatov:2022enf,Borah:2025crf, Rehman:2025hfd, Shaymatov:2023rgb, Stuchlik:2015sno, Banerjee:2022ffu}.

It is worth noting that the Kerr–Bertotti–Robinson spacetime solution is an exact solution of the Einstein–Maxwell equations describing a rotating black hole immersed in a uniform electromagnetic field \cite{rfgvybz5}. This is obtained by introducing the external electromagnetic field in the Kerr–Newman family. This was determined by astrophysically because BHs in the universe are typically surrounded by a magnetic field, which serves as a foundation for explaining the high-energy activity of quasars and galactic nuclei. Recent studies demonstrate that magnetized spacetimes, where magnetic fields belong to the geometry, can substantially influence geodesic dynamics, black hole shadow properties, and energy extraction mechanisms (see for example \cite{Wang2025vsx,Zeng:2025KRB,Vachher2025JCAP,Wang2025bjf,Siahaan2025KRB,Mirkhaydarov2026MPP,Liu2025wwq,Xamidov:2026kqs}). In this work, we focus on a rotating Kerr black hole immersed in a uniform Bertotti–Robinson magnetic field, which provides a suitable framework for studying QPOs. We analyze the motion of test particles by deriving the geodesic equations and demonstrate that the magnetic field parameter can significantly modify the orbital, radial, and vertical epicyclic frequencies, as well as the associated resonance conditions. We then employ these fundamental frequencies to construct theoretical QPO models based on parametric and forced resonance mechanisms. Using observational data from black hole X-ray binaries, including GRO J1655–40, XTE J1550–564, XTE J1859+226, GRS 1915+105, H1743–322, M82 X–1, and Sgr~A$^{*}$, we constrain the model parameters using Bayesian inference and Markov Chain Monte Carlo (MCMC) methods. Finally, we investigate the impact of the magnetic field parameter on the location of the innermost stable circular orbit and on the radiative properties of thin accretion disks. Our findings suggest that magnetized environments can lead to observable deviations from the Kerr black hole scenario in X-ray observations. In particular, QPO observations can provide constraints on the magnetic field parameter and the underlying spacetime properties.

These solutions, initially derived by Ernst and Wild using the Harrison transformation, are known as Schwarzschild–Melvin, Kerr–Melvin, and Kerr–Newman–Melvin spacetimes \cite{Ernst:1976mzr, Ernst:1976bsr}. They represent BHs interacting with an external magnetic field from the Bonner–Melvin universe \cite{griffiths2009exact,bonnor1954static}. Additional intriguing extensions have been identified, with the most recent ones discussed in \cite{di2025kerr}. These BHs have been extensively utilized for analyzing different physical phenomena. For detailed reviews of the main findings, refer to \cite{Aliev:1989wx, Gibbons:2013dna}. Specifically, researchers have studied the interaction between the external field and the BH, resulting in the identification of the Meissner effect. This effect reveals that BHs expel magnetic fields from their horizons when they reach extremal states (see for example \cite{Wald:1974np,Bicak:1980du,Bicak:2015lxa,Gurlebeck:2018smy}). There are some drawbacks associated with Melvin-like BH spacetimes; therefore, to prevent them, one assumes completely realistic global models. It is worth noting that the magnetic field reduces far from the BH, yet geodesics are unable to reach infinity. They even exhibit chaotic behavior \cite{karas1992chaotic}, and their ergoregions extend to infinity \cite{gibbons2013ergoregions,Shaymatov22PhRvD.106b4039S}. Furthermore, researchers investigate in \cite{Pravda:2005uv} that these spacetimes are of algebraic type I, representing that such ``standing'' BHs may be radiative.

The following is the structure of the article. Section II focuses on computing the fundamental QPO frequencies from the Euler-Lagrange equations of motion for massive particles in the vicinity of the considered BH. In Sec .~III, we discuss the PR and FR models. In Sec.~V, we analyze the observational data from QPO X-ray binaries and perform the MCMC analysis. Section VI presents the constraints on the BH parameters, including the best-fit values obtained through the MCMC analysis, and also computes the ISCO orbit and the corresponding physical quantities at the ISCO. In Sec.~VII, we examine the energy flux and temperature of the accretion disk near the considered BH. Finally, we summarize our main findings.

\section{Mathematical framework for Kerr BH in a uniform Bertotti-Robinson magnetic field and associated QPO frequencies}

The line element in Boyer-Lindquist coordinates for the Kerr BH in a uniform Bertotti-Robinson magnetic field \cite{rfgvybz5} can be presented as:  
\begin{eqnarray}
ds^{2}&&=\frac{1}{\Omega^{2}}\Big[
-\frac{Q}{\rho^{2}}\left(dt-a\sin^{2}\theta\, d\varphi\right)^{2}
+\frac{\rho^{2}}{Q}\,dr^{2}
\\&& \nonumber +\frac{\rho^{2}}{P}\,d\theta^{2}
 +\frac{P}{\rho^{2}}\sin^{2}\theta
(a\,dt-(r^{2}+a^{2})\,d\varphi)^{2}
\Big],   
\end{eqnarray}
where
\begin{eqnarray}
  \rho^{2}&=&r^{2}+a^{2}\cos^{2}\theta,\\
P&=&1+B^{2}\left(m^{2}\frac{I_{2}}{I_{1}^{2}}-a^{2}\right)\cos^{2}\theta,\\
Q&=&(1+B^{2}r^{2})\,\Delta,\\
\Omega^{2}&=&(1+B^{2}r^{2})-B^{2}\Delta\cos^{2}\theta,\\
\Delta&=&\left(1-B^{2}m^{2}\frac{I_{2}}{I_{1}^{2}}\right)r^{2}-2m\frac{I_{2}}{I_{1}}\,r+a^{2},\\
I_{1}&=&1-\frac{1}{2}B^{2}a^{2},\\
I_{2}&=&1-B^{2}a^{2}.  
\end{eqnarray}
Here $m$, $a=J/m$, and $Bm=b$ represent the BH mass, the angular momentum per unit mass (with total angular momentum $J$), and the dimensionless magnetic-field parameter, respectively. This section focuses on the QPOs in the vicinity of the Kerr BH in a uniform Bertotti–Robinson magnetic field. In order to investigate QPOs around the Kerr BH in a uniform Bertotti–Robinson magnetic field, we examine the geodesic motion of the particles and compute the fundamental frequencies, which define their movement within the strong gravitational field. 
The investigation starts with the Lagrangian of the particle
\begin{equation}
\mathcal{L} = \frac{1}{2} g_{\mu\nu} \frac{dx^\mu}{d\lambda} \frac{dx^\nu}{d\lambda},\label{za2}
\end{equation}
where $\lambda$ represents the worldline affine parameter of the particle. For massive particles $\mathcal{L}<0$ and for massless particles, $\mathcal{L}=0$. The associated generalized momentum can be expressed as
\begin{equation}
p_\mu = \frac{\partial \mathcal{L}}{\partial \dot{x}^\mu} = g_{\mu\nu} \dot{x}^\nu. \label{a1}
\end{equation}
By using Eq.~(\ref{a1}), we obtain
\begin{eqnarray}
p_t &=& g_{tt} \dot{t} + g_{t\phi} \dot{\phi} = -\tilde{E}, \label{a2}\\
p_\phi &=& g_{t\phi} \dot{t} + g_{\phi\phi} \dot{\phi} = \tilde{L}, \label{a3}\\
p_r &=& g_{rr} \dot{r}, \\
p_\theta &=& g_{\theta\theta} \dot{\theta}. \label{a4}
\end{eqnarray}
In the above equations, the overdot represents differentiation with respect to the affine parameter $\lambda$, while $\tilde{E}$ and $\tilde{L}$ indicate the conserved energy and conserved angular momentum, respectively. By solving these equations, we have
\begin{eqnarray}
\dot{t} &=& \frac{g_{\phi\phi} \tilde{E} + g_{t\phi} \tilde{L}}{g_{t\phi}^2 - g_{tt} g_{\phi\phi}}, \label{a5}\\
\dot{\phi} &=& \frac{\tilde{E} g_{t\phi} + g_{tt} \tilde{L}}{g_{tt} g_{\phi\phi} - g_{t\phi}^2}. \label{a6}
\end{eqnarray}
By Employing condition, $g_{\mu \nu} \, \dot{x}^{\mu} \, \dot{x}^{\nu} = -1$ along with Eqs.~(\ref{a5}) and (\ref{a6}), one can obtain 
\begin{equation}
g_{rr} \, \dot{r}^{2} + g_{\theta\theta} \, \dot{\theta}^{2} = -1 - g_{tt} \, \dot{t}^{2} - g_{\phi\phi} \, \dot{\phi}^{2} - 2 g_{t\phi} \, \dot{t} \, \dot{\phi}. \label{a7}
\end{equation}
 We restricted the motion of the particles to a plane, i.e., $\theta = \pi/2$ and $\dot{\theta} = 0$. From Eqs.~(\ref{a5})--(\ref{a7}), we have
\begin{equation}
\dot{r}^{2} = V_{\text{eff}}(M, r, \tilde{L},\tilde{E}) =
\frac{\tilde{E}^{2} g_{\phi\phi} + 2 \tilde{E} \tilde{L} g_{t\phi} + \tilde{L}^{2} g_{tt}}{g_{t\phi}^{2} - g_{tt} g_{\phi\phi}} - 1. \label{a8}
\end{equation}
In the above expression, the $V_{\text{eff}}(r, M, \tilde{E}, \tilde{L})$ indicates the effective potential governing the motion of the test particle characterized by the specific energy $\tilde{E}$ and specific angular momentum $\tilde{L}$. It is worth noting that stable circular orbits satisfy the conditions $\dot{r}=0$ and $dV_{\text{eff}}/dr=0$. From the given conditions, we obtain the expressions for $\tilde{E}$, and $\tilde{L}$ 
\begin{eqnarray}
\tilde{E} = \frac{-g_{tt} + g_{t\phi}\Omega_\phi}{\sqrt{ -g_{tt} - 2g_{t\phi}\Omega_\phi - g_{\phi\phi}\Omega_\phi^2 }}, \label{a9} \\
\tilde{L} = \frac{g_{t\phi} + g_{\phi\phi}\Omega_\phi}{\sqrt{ -g_{tt} - 2g_{t\phi}\Omega_\phi - g_{\phi\phi}\Omega_\phi^2 }}. \label{a10}
\end{eqnarray}
Here, $\Omega_{\phi}$ is the angular velocity of particles on circular orbits, obtained by
\begin{equation}
\Omega_\phi = \frac{ -\partial_r g_{t\phi} \pm \sqrt{ (\partial_r g_{t\phi})^2 - (\partial_r g_{tt})(\partial_r g_{\phi\phi}) } }{ \partial_r g_{\phi\phi} }.\label{a11}
\end{equation}
Here, we investigate QPOs by computing the orbital frequency $\nu_\phi$, the vertical epicyclic frequency $\nu_{\theta}$, and the radial epicyclic frequency $\nu_r$, with the orbital frequency given by
\begin{eqnarray}
    \nu_\phi=\frac{\Omega_\phi}{2\pi}. \label{orbital}
\end{eqnarray}
By considering small perturbations near the circular orbit in the equatorial plane, the radial and vertical epicyclic frequencies are obtained by
\begin{eqnarray}
\theta(t) = \frac{\pi}{2} + \delta \theta(t), \quad r(t) = r_0 + \delta r(t), \label{a14}
\end{eqnarray}
where, $\delta r(t)$ and $\delta \theta(t)$ correspond to small perturbations that are responsible for the subsequent expression
\begin{eqnarray}
\frac{d^2 \delta \theta(t)}{dt^2} + \Omega_\theta^2 \delta \theta(t) = 0,\label{a15}\\
\frac{d^2 \delta r(t)}{dt^2} + \Omega_r^2 \delta r(t) = 0, \label{a16}
\end{eqnarray}
where
\begin{eqnarray}
\Omega_\theta^2 = -\frac{1}{2 g_{\theta\theta} \dot{t}^2} \left. \frac{\partial^2 V_{\text{eff}}}{\partial \theta^2} \right|_{\theta = \frac{\pi}{2}}, \label{a17}\\
\Omega_r^2 = -\frac{1}{2 g_{rr} \dot{t}^2} \left. \frac{\partial^2 V_{\text{eff}}}{\partial r^2} \right|_{\theta = \frac{\pi}{2}}, \label{a18}
\end{eqnarray}
From Eqs.~(\ref{a17}) and (\ref{a18}), we computed the vertical and radial epicyclic frequencies 
\begin{eqnarray}
\nu_\theta &= \frac{\Omega_\theta}{2\pi}\, \, \mbox{~and~}\,\,
\nu_r &= \frac{\Omega_r}{2\pi}\, . \label{radial}
\end{eqnarray}
From  Eqs. (\ref{orbital}) and (\ref{radial}), one can compute the exact solutions of the fundamental frequencies for the BH under consideration
\section{The frequency of a prescription for quasiperiodic oscillations.}
Several theoretical frameworks have been proposed to explain the QPO phenomenon. This section focuses on the PR and FR model. The PR model attributes QPOs to nonlinear interactions between the radial and vertical oscillatory modes of particle motion, whereas the FR model interprets the oscillations as the result of external perturbations or a driven accretion disk
\subsection{Parametric Resonance Mode}
Twin-peak high-frequency QPOs observed in BH and neutron star systems often display a characteristic $3:2$ frequency ratio. This recurrent pattern has led to the suggestion that these oscillations may arise from nonlinear resonant interactions between different modes of motion in the accretion disk \cite{Kluzniak:2002bb, Abramowicz:2001bi, Abramowicz:2003xy, Rebusco:2004ba,  Abramowicz:2001bi, Abramowicz:2004je}. In a simple description, small deviations from circular equatorial geodesic motion can be treated as harmonic oscillations in the vertical and radial directions. These motions have been described by the radial and vertical epicyclic frequencies, denoted by $\nu_r$ and $\nu_\theta$.

Within the parametric resonance (PR) framework, radial oscillations can act as a driving mechanism that excites vertical oscillations via nonlinear coupling. In the thin disks, the amplitude of radial perturbations is usually larger than that of vertical perturbations ($\delta r > \delta \theta$), which allows radial motion to parametrically amplify vertical oscillations once a resonance condition is satisfied. The resonance condition is achieved when the ratio of the epicyclic frequencies satisfies
\[
\frac{\nu_r}{\nu_\theta} = \frac{2}{n},
\]
here, $n$ denotes a positive integer. For rotating BHs, the relation $\nu_\theta > \nu_r$ is generally satisfied, and the resonance becomes particularly significant for $n=3$. This condition naturally produces the commonly observed $3:2$ ratio of twin-peak QPO frequencies. In this interpretation, the upper and lower observed frequencies are associated with the vertical and radial epicyclic frequencies, respectively, such that $\nu_U = \nu_\theta$ and $\nu_L = \nu_r$.
\subsection {Forced Resonance Model}
In many astrophysical phenomena, the thin Keplerian disk \cite{Kluzniak:2002bb, Abramowicz:2001bi, 2001AcPPB..32.3605K, 2005A&A...436....1T} does not adequately describe the accretion flow because viscosity, pressure, and magnetic stresses can significantly modify the disk dynamics. These effects can lead to a nonlinear relationship between the perturbations $\delta \theta$ and $\delta r$, in addition to the previously discussed parametric resonance. Furthermore, numerical simulations have shown that radial oscillations can excite vertical oscillations through pressure coupling~\cite{Abramowicz:2001bi, Lee:2004bp}. The nonlinear coupling between $\delta \theta$, and $\delta r$ is frequently represented utilizing a mathematical ansatz
\begin{eqnarray}
{\delta \ddot\theta} + \delta \theta \omega_{\theta}^{2} = -\omega_{\theta}^{2}\,\delta r\,\delta \theta + \mathcal{F}_{\theta}(\delta \theta), \label{fmodel}
\end{eqnarray}
here, $\mathcal{F}_{\theta}$ and $\delta r=Acos(\omega_{r}t)$ represents the nonlinear terms in $\delta \theta$. From Eq. \ref{fmodel} we have
\begin{equation}
\frac{\nu_{\theta}}{\nu_{r}} = \frac{m}{n}, \qquad \text{where $m$ and $n$ are natural numbers.}
\end{equation}
For the forced resonance model $m:n = 3:1$, the expressions for the  upper and lower frequencies are obtained as
\begin{eqnarray}
\nu_{L} &=& \nu_\theta - \nu_r, \\
\nu_{U} &=& \nu_\theta .
\label{FR}
\end{eqnarray}

\section{Observational Analysis}

\begin{table*}[t]
\begin{ruledtabular}
\caption{For the observational analysis, we consider the QPO X-ray binaries listed in the table.}
\label{tab: I}
\begin{tabular}{c|c|c|c|c|c|c|c}
& GRO J1655--40 & XTE J1550--564 & XTE J1859+226 & GRS 1915+105 & H1743--322 & M82\,X{-1} & Sgr A$^{*}$ \\
\hline
$M~(M_{\odot})$
& $5.4 \pm 0.3$~\cite{Motta:2013wga}
& $9.1 \pm 0.61$~\cite{Remillard:2002cy,Orosz:2011ki}
& $7.85 \pm 0.46$~\cite{Motta:2022rku}
& $12.4^{+2.0}_{-1.8}$~\cite{Remillard:2006fc}
& $\gtrsim 9.29$~\cite{Ingram:2014ara}
& $415 \pm 63$~\cite{Pasham2014}
& $(3.5\text{--}4.9)\cdot 10^{6}$~\cite{Ghez:2008ms, Gillessen:2008qv}
\\
$\nu_{U}\,(\mathrm{Hz})$
& $441 \pm 2$~\cite{Motta:2013wga}
& $276 \pm 3$~\cite{Remillard:2002cy}
& $227.5^{+2.1}_{-2.4}$~\cite{Motta:2022rku}
& $168 \pm 3$~\cite{Remillard:2006fc}
& $240 \pm 3$~\cite{Ingram:2014ara}
& $5.07 \pm 0.06$~\cite{Pasham2014}
& $(1.45 \pm 0.16)\times 10^{-3}$~\cite{Stuchlik:2008fy}
\\
$\nu_{L}\,(\mathrm{Hz})$
& $298 \pm 4$~\cite{Motta:2013wga}
& $184 \pm 5$~\cite{Remillard:2002cy}
& $128.6^{+1.6}_{-1.8}$~\cite{Motta:2022rku}
& $113 \pm 5$~\cite{Remillard:2006fc}
& $165^{+9}_{-5}$~\cite{Ingram:2014ara}
& $3.32 \pm 0.06$~\cite{Pasham2014}
& $(0.89 \pm 0.04)\times 10^{-3}$~\cite{Stuchlik:2008fy}
\\
$\nu_{C}\,(\mathrm{Hz})$
& $17.3 \pm 0.1$~\cite{Motta:2013wga}
& --
& $3.65 \pm 0.01$~\cite{Motta:2022rku}
& --
& $9.44 \pm 0.02$~\cite{Ingram:2014ara}
& --
& --
\end{tabular}
\end{ruledtabular}
\end{table*}

\begin{table}
\begin{ruledtabular}
\caption{Prior ranges for the model parameters are specified by adopting uniform $(\mathcal{U})$ and Gaussian $(\mathcal{N}(\mu,\sigma))$ distributions, constrained using the observational data set.}
\label{tab:placeholder}
\begin{tabular}{cc}
Parameters & Prior Range\\
\hline
\(m(m_\odot)\)& $ \mathcal{U} [1,10^7]$\\  
\(a/m\) & \( \mathcal{N}(0.3, 0.05)\) \\
\(r/m\) & \(\mathcal{N} (6.8, 0.5)\) \\
\(b\) & $ \mathcal{U} [0,0.1]$
\end{tabular}
\end{ruledtabular}
\end{table}

This section focuses on constraining the considered BH parameters, namely the mass $m$, orbital radius $r/m$, dimensionless spin parameter $a/m$, and dimensionless magnetic field parameter $b$, by employing the observational data provided in Table~\ref{1a}. In this table, we consider seven observational samples with BH masses (in solar units) and the corresponding lower, upper, and nodal frequencies. Missing data in the table are represented by a dash. 
       
To constrain the BH parameters, we consider the PR and FR models and apply the observational data corresponding to the doublet $\{\nu_{U}, \nu_{L}\}$. Bayes' theorem is then used to compute the posterior distribution of the model parameters:
\begin{equation}
P(\boldsymbol{\theta} | D, H) = \frac{P(D | \boldsymbol{\theta}, H) \, P(\boldsymbol{\theta} | H)}{P(D | H)} \, ,
\end{equation}
in the above expression, the symbols $D$, $H$, and $\boldsymbol{\theta}$ represent the data vector, the model hypothesis, and the parameter vector, respectively. We compute the logarithm of the likelihood, $\mathcal{L} \equiv P(D \mid \boldsymbol{\theta}, H)$, using the $\chi^{2}$, defined by:
\begin{equation}
\ln	\mathcal{L} =\left(-\frac{1}{2} \chi^2 \right) \,,
\end{equation}
where $\chi^2$ is
\begin{equation}
\chi^2 = \sum_{i=1}^{N} \left( \frac{D_{i, \mathrm{Obs}} - D_{i, \mathrm{Model}}}{\sigma_i} \right)^2.
\end{equation}
Here $\sigma_i$ represents uncertainty corresponding to every observational measurement. This analysis is performed by assuming uniform (flat) priors for the parameters, together with broad Gaussian priors for parameters whose ranges are uncertain but approximately known \cite{Padilla:2019mgi,Rehman:2025knv}.
The likelihood is computed using a Python-based pipeline. The \emph{dynamic nested sampling} algorithm \texttt{dynesty} is efficient for investigating multimodal or degenerate posterior distributions \cite{Higson:2018cwj}. The obtained posterior samples are examined using the \texttt{GetDist} package to derive marginalized constraints on individual parameters and to generate one- and two-dimensional posterior distributions \cite{Lewis:2019xzd}. Parameter estimates are presented at the 68\% confidence level (CL), unless stated otherwise.

In this study, we apply Gaussian priors with large dispersion to the $r/m$ and $a/m$, while uniform priors are adopted for the BH mass $m$ and the dimensionless magnetic field parameter $b$. The prior ranges for the BH parameters are provided in Table~\ref{tab:placeholder}.

\section{Results and Discussion}

\subsection{Case II: Parametric Resonance Model (PR)}

\begin{figure*}
\centering
\includegraphics[scale=0.3]{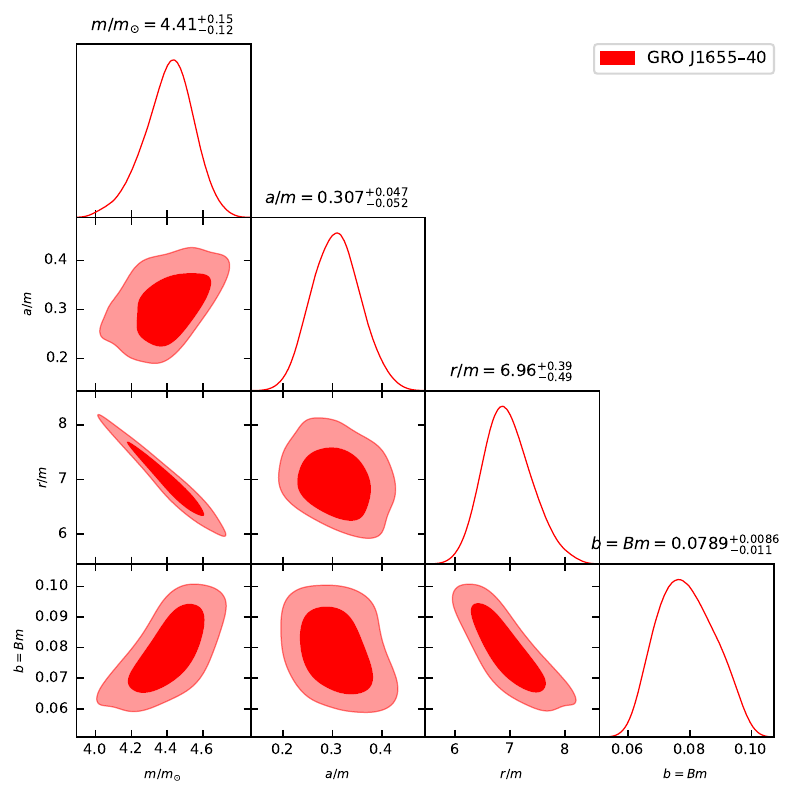}
\includegraphics[scale=0.3]{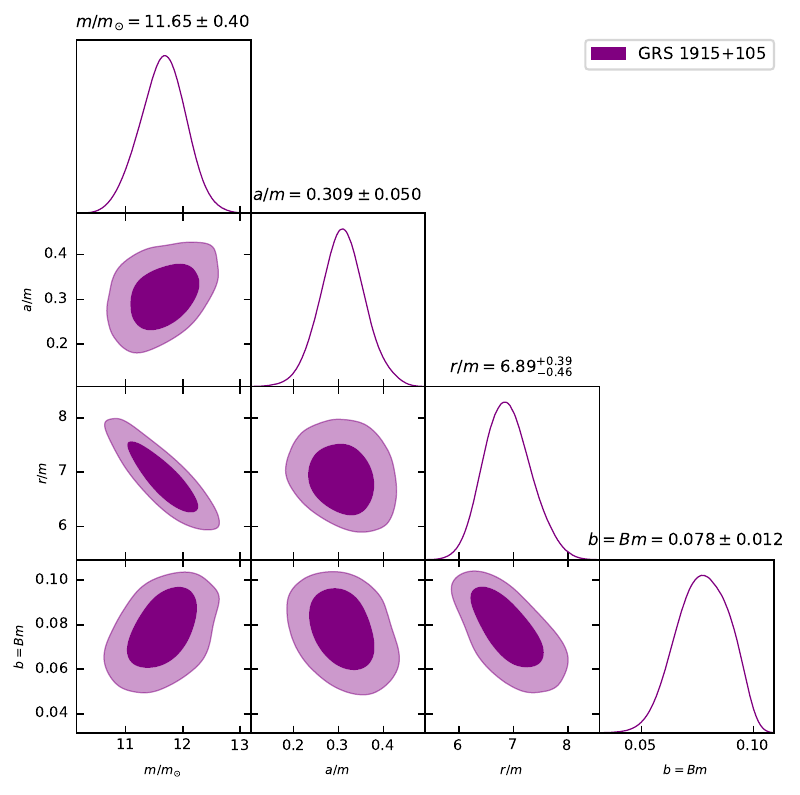}
\includegraphics[scale=0.3]{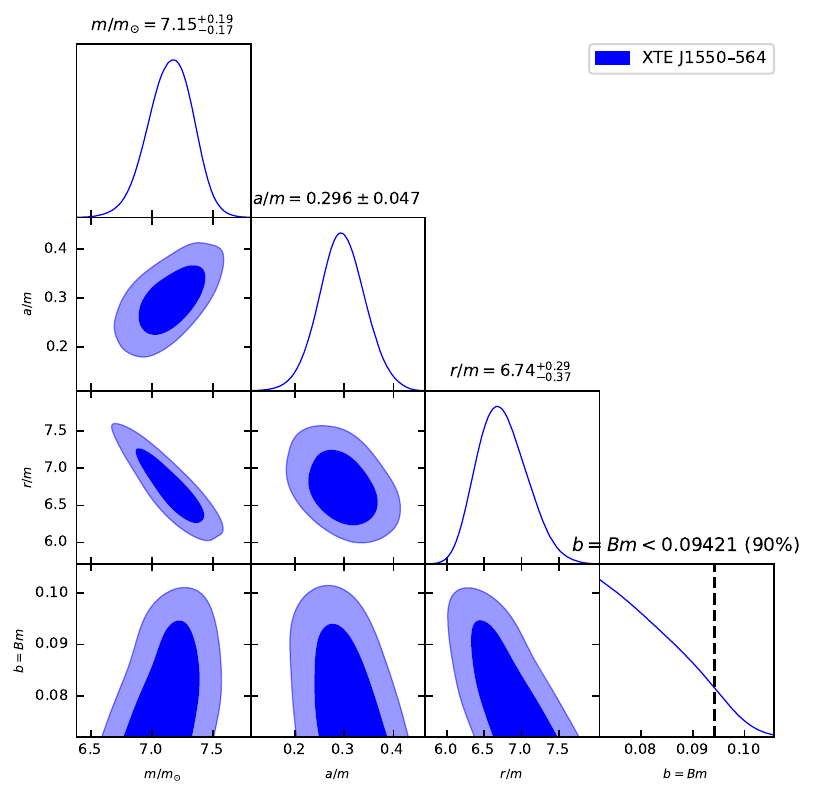}
\includegraphics[scale=0.3]{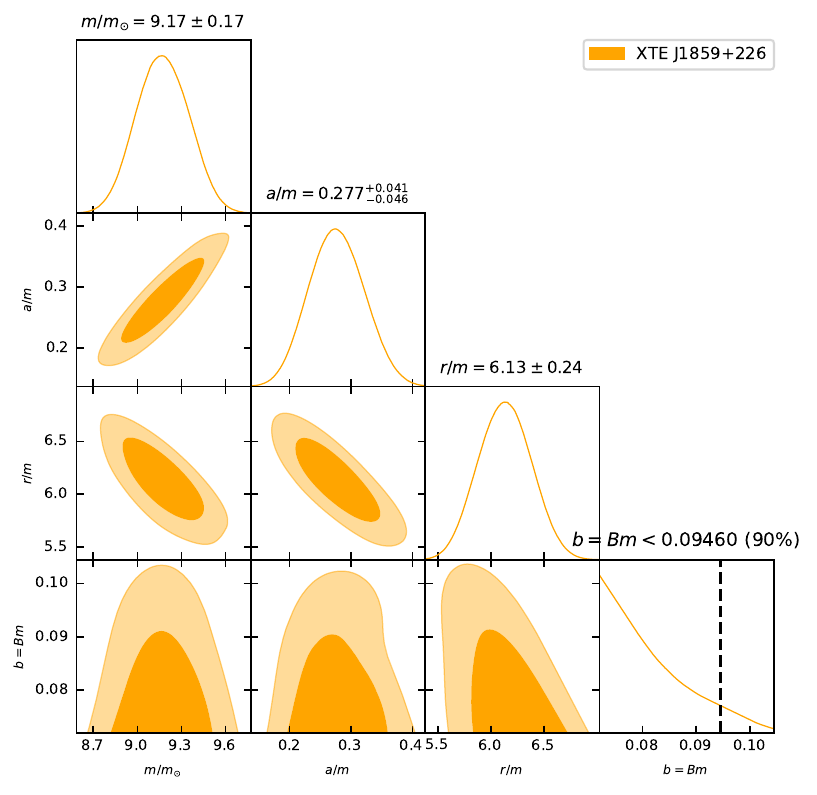}
\includegraphics[scale=0.3]{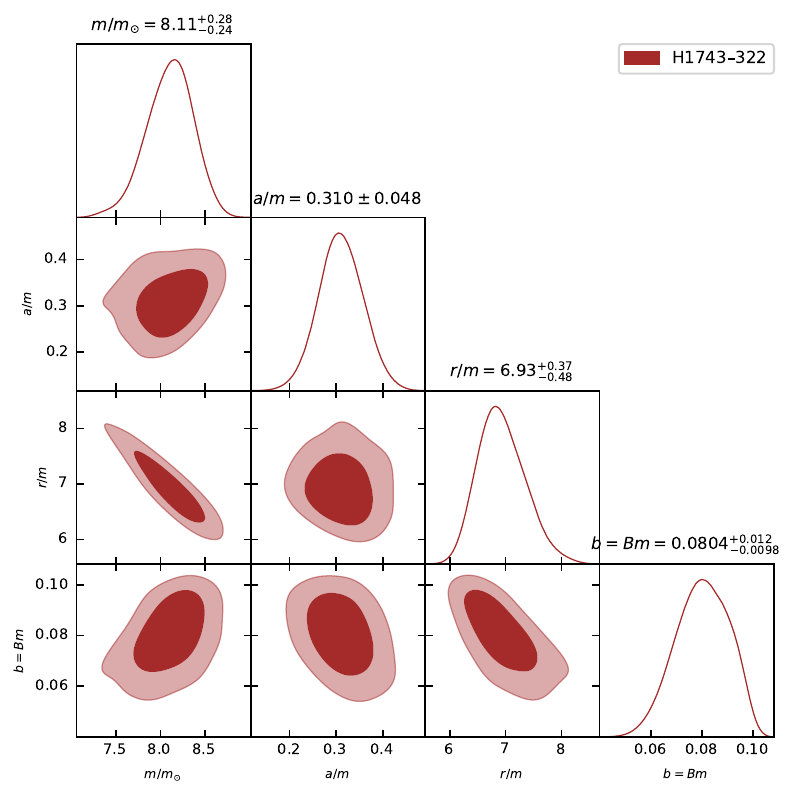}
\includegraphics[scale=0.3]{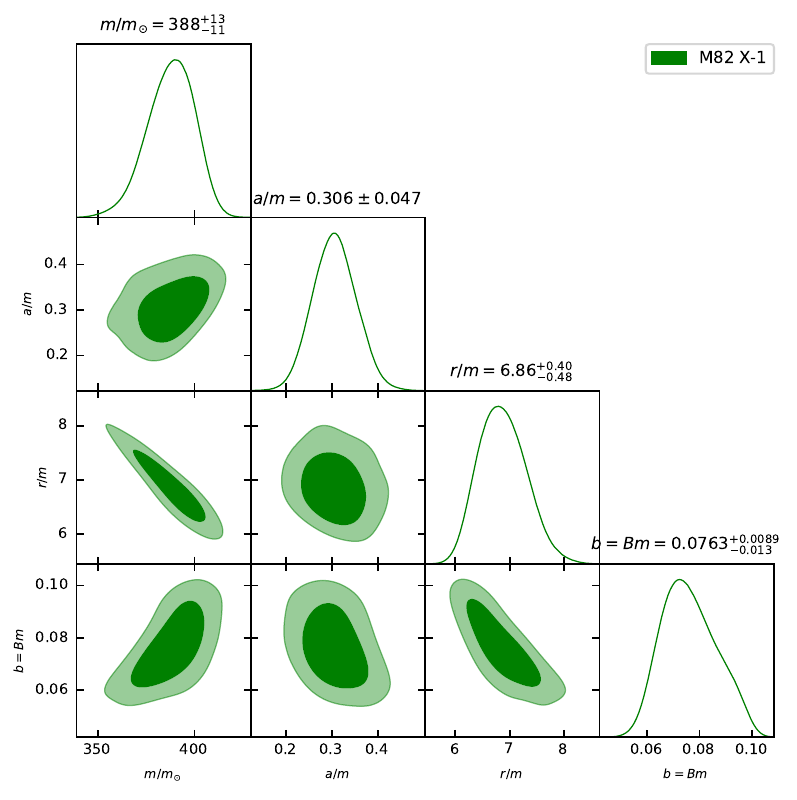}
\includegraphics[scale=0.3]{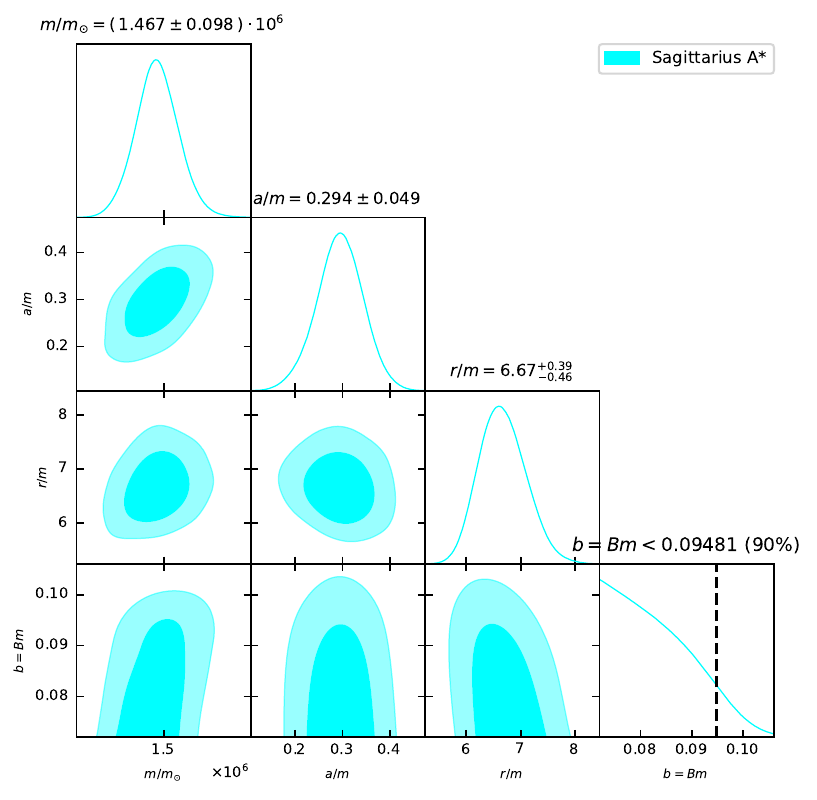}
\caption{For several observed QPO X-ray binaries, posterior distributions of the BH parameters $(m, a/m, r/m, b)$ are obtained from the MCMC analysis within the framework of the PR model. The off-diagonal panels display the corresponding two-dimensional confidence contours, while the diagonal panels show the marginalized one-dimensional posterior distributions of each parameter.}
\label{1a}
\end{figure*}

In this subsection, we study the four-dimensional parameter space $(m, a/m, r/m, b)$ of the Kerr-Bertotti-Robinson BH applying MCMC analysis within the framework of the PR model. Figure~\ref{1a} shows the posterior distributions obtained for the PR model, while the obtained best-fit values are presented in Table~\ref{tab:bestfit}. In this scenario, the BH masses are consistent with the observational data given in Table~\ref{tab: I}, with only mild deviations. The spin parameter $a/m$ lies in the range $0.294-0.310$, indicating that the BH is moderately rotating. The resonance radius $r/m$ lies within the range $6.13-6.96$, which is close to the inner region of the accretion disk. This suggests that the mechanism responsible for the observed QPOs occurs in the inner region of the accretion disk. Furthermore, the posterior distributions indicate that the spin parameter and the resonance radius are positively correlated because an increase in the spin parameter shifts the resonance radius outward. This occurs because the epicyclic frequencies depend on the spacetime geometry. As a consequence, a change in the spin parameter can be compensated by a corresponding change in the resonance radius while still reproducing the observed QPO frequencies.

In this study, an important aspect is the constraint on the external magnetic field parameter $b$. The posterior analysis shows that this parameter is small but non-zero for the Kerr-Bertotti-Robinson spacetime. It is worth noting that for several X-ray binaries, namely GRS 1915+105, GRO J1655-40, M82 X-1, and  H1743-322, our analysis indicates non-zero values of the magnetic field parameter constrained at the $68\%$ confidence level (C.L.). The obtained values are $b =0.078\pm 0.012$, $b = 0.0789_{-0.011}^{+0.0086}$, $b = 0.0763^{+0.0089}_{-0.013}$, and $b =  0.0804^{+0.012}_{0.0098}$, respectively. These results suggest the presence of magnetic corrections associated with the Kerr-Bertotti-Robinson spacetime for these systems. For the remaining sources, however, only an upper bound at the $90\%$ C.L. is obtained, indicating that the data are consistent with a Kerr-like spacetime within the current observational uncertainties. 

From this behavior, it is evident that the sensitivity of the QPO frequencies to the magnetic field introduced by the spacetime geometry is relatively weak. Consequently, the dominant contribution to the epicyclic frequencies arises from the gravitational parameters, namely the mass $m$, the spin parameter $a/m$, and the resonance radius $r/m$, while the magnetic field parameter $b$ contributes only as a subleading correction and may therefore be weakly constrained by the available data.

In this study, one of the important aspects is constraining the external magnetic field parameter $b$, and the posterior analysis shows that this parameter is small but non-zero for the Kerr-Bertotti-Robinson spacetime. For several X-ray binaries, namely  GRS 1915+105, GRO J1655-40, M82 X-1, and H1743-322, the magnetic field parameter is constrained at the $68\%$ confidence level (C.L.), while for the remaining cases we obtain only an upper bound at the $90\%$ C.L. From this behavior, it is evident that the sensitivity of the QPO frequencies to the magnetic field introduced by the spacetime geometry is relatively weak. Therefore, the dominant contribution to the epicyclic frequencies arises from the gravitational parameters, namely the mass $m$, the spin parameter $a/m$, and the resonance radius $r/m$, while the magnetic field parameter $b$ contributes as a subleading correction and may therefore be weakly constrained by the available data. 

However, the small inferred values of the magnetic field parameter $b$ modify the effective potential governing test particle motion, which shifts the epicyclic frequencies and influences the orbital dynamics of particles in the accretion disk. This modification changes the radial location where the resonance condition between oscillation modes is satisfied. Thus, our results show that QPO observations can place meaningful constraints on the magnetic coupling parameter associated with the electromagnetic field of the Kerr-Bertotti-Robinson spacetime.

\subsection{Case III: Forced Resonance Model (FR)}
\begin{figure*}
\centering
\includegraphics[scale=0.3]{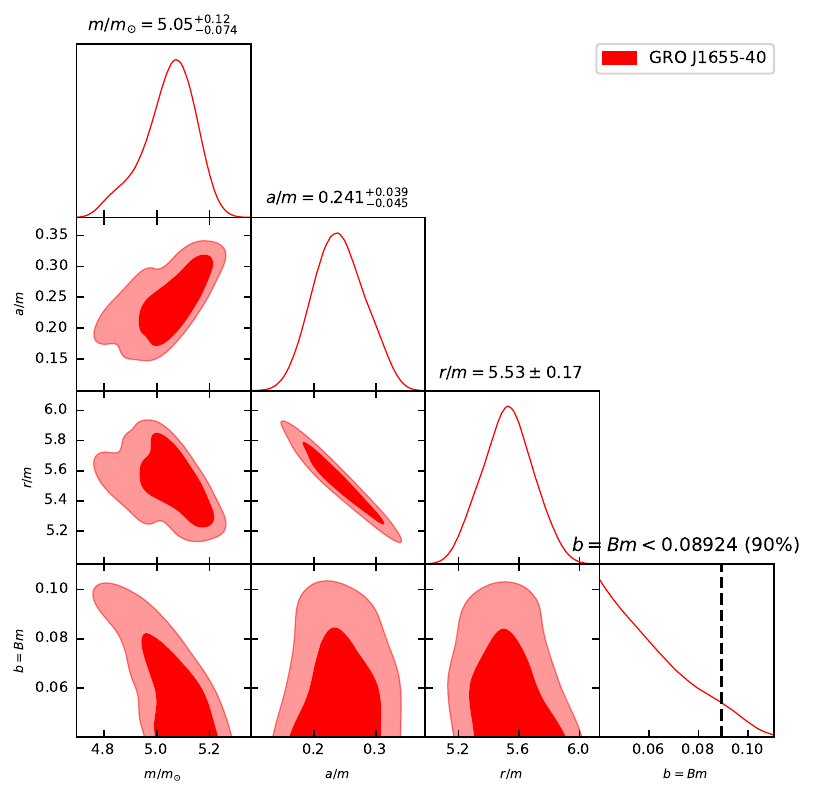}
\includegraphics[scale=0.3]{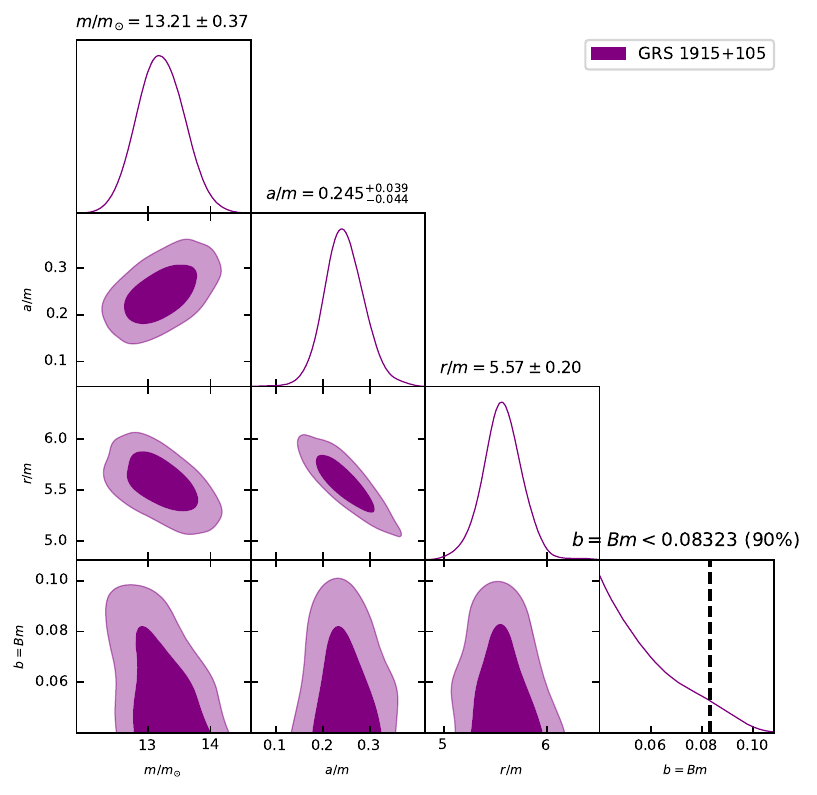}
\includegraphics[scale=0.3]{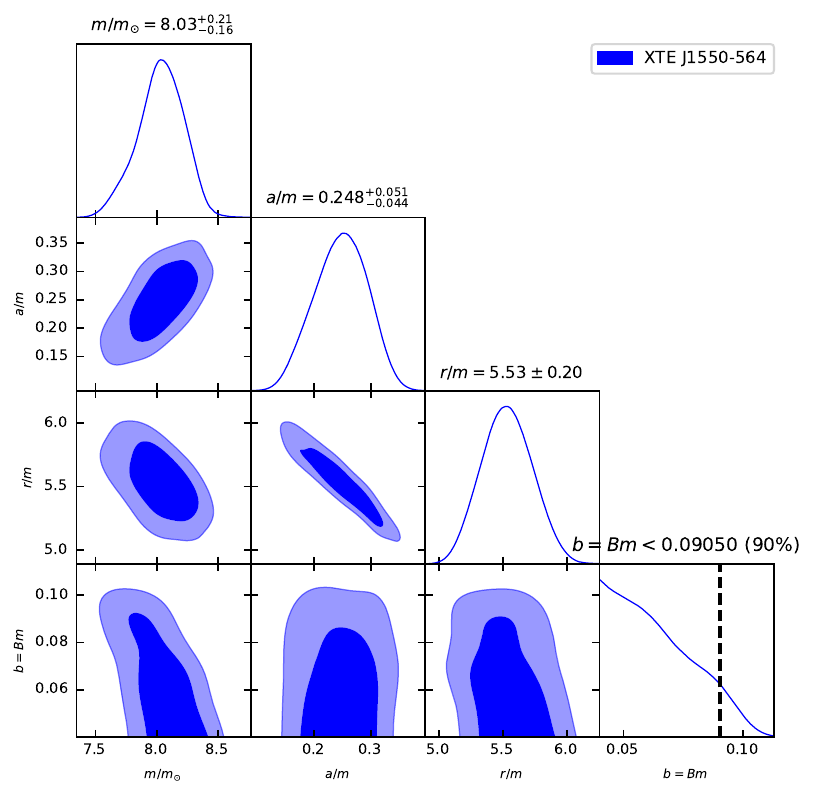}
\includegraphics[scale=0.3]{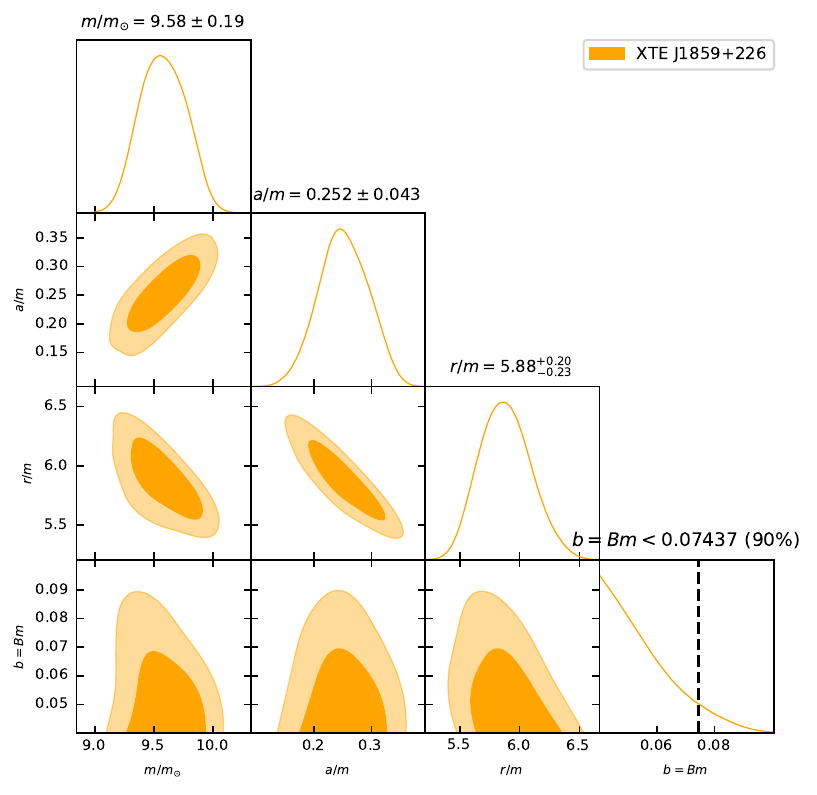}
\includegraphics[scale=0.3]{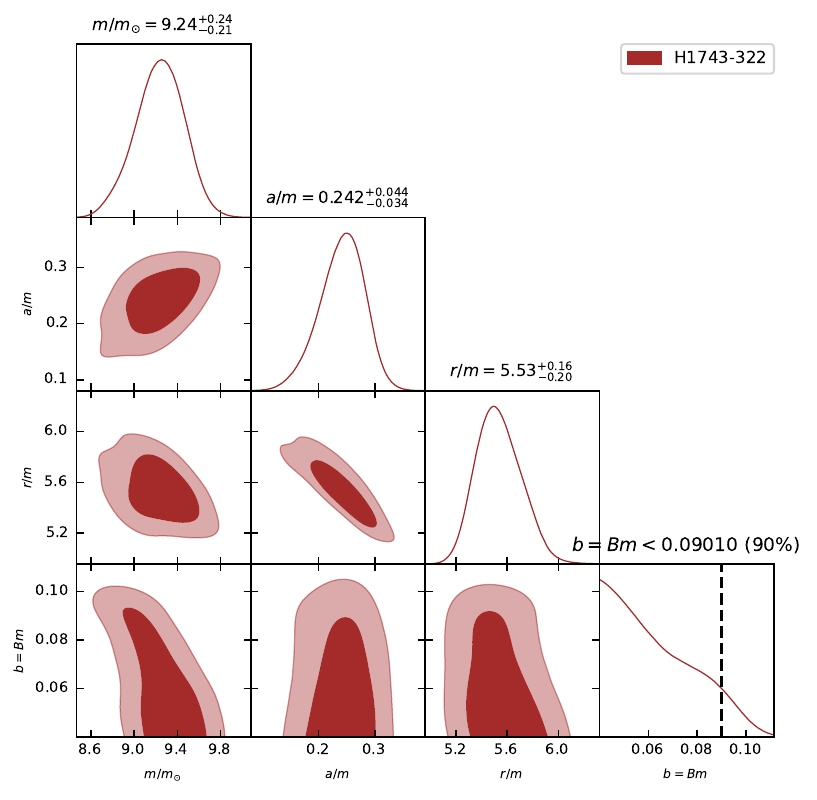}
\includegraphics[scale=0.3]{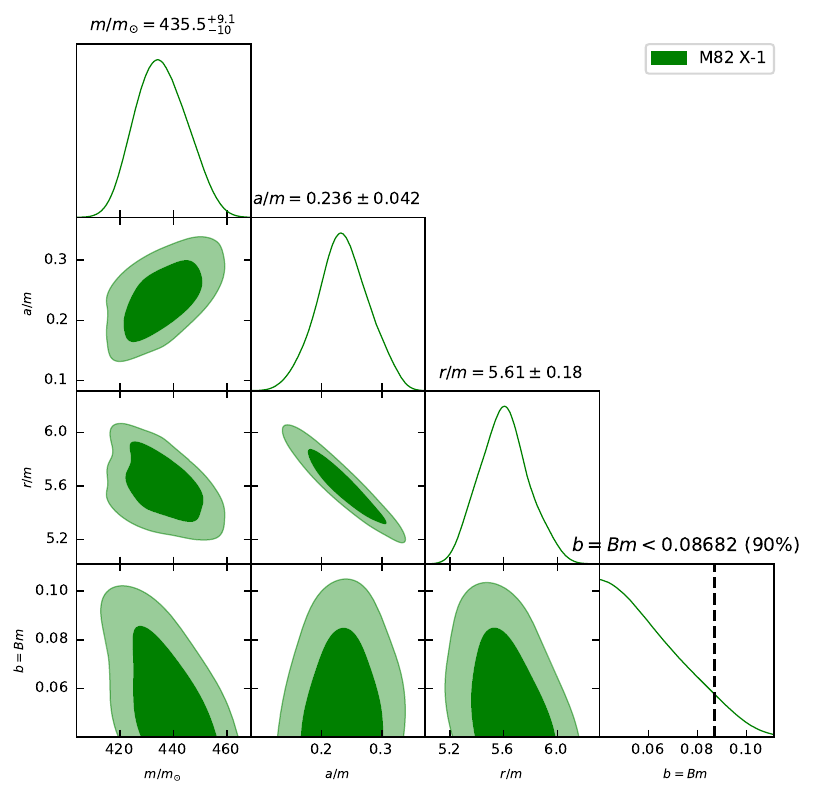}
\includegraphics[scale=0.3]{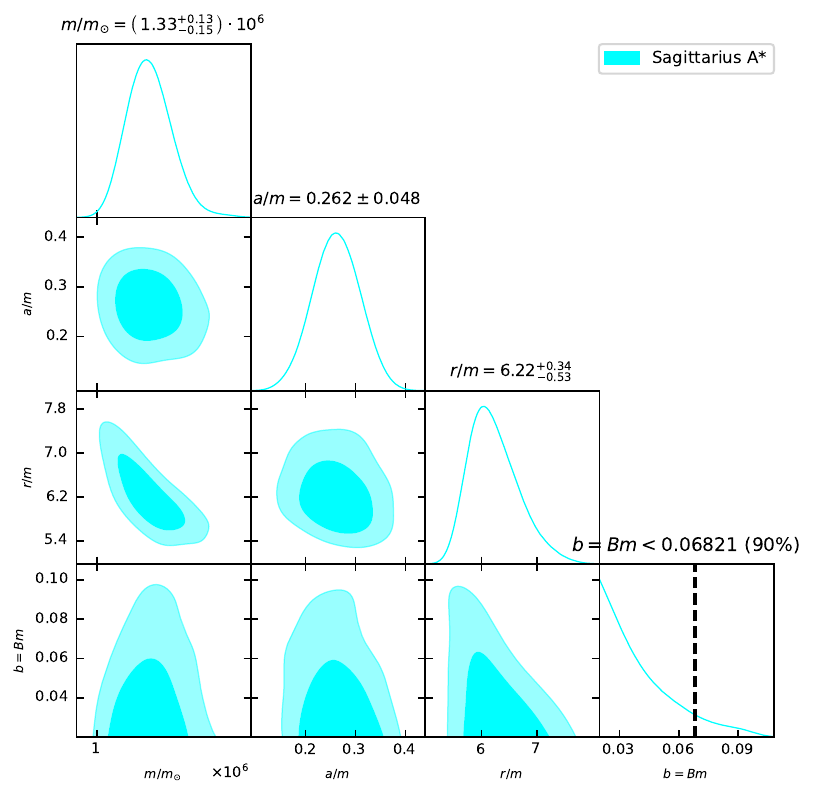}
\caption{The MCMC analysis within the FR model yields posterior distributions of the BH parameters $(m, a/m, r/m, b)$ for the considered QPO X-ray binaries. The magnetic field parameter $b$ is constrained by an upper limit at the $90\%$ confidence level, whereas the mass $m$, spin parameter $a/m$, and resonance radius $r/m$ are constrained at the $68\%$ confidence level.}
\label{2a}
\end{figure*}

This subsection focuses on investigating the four-dimensional parameter space $(m,a/m,r/m,b)$ of the considered spacetime by employing an MCMC analysis within the framework of the FR model. The posterior distributions of the BH parameters for the considered X-ray binaries are presented in Fig.~\ref{2a}, and the associated best-fit values are given in Table~\ref{tab:bestfit}. 

In this case, the BH mass, spin parameter, and resonance radius are constrained at the $68\%$ confidence interval, while the magnetic field parameter is constrained at the $90\%$ confidence interval. The inferred BH masses are consistent with the observational data provided in Table~\ref{1a}. The spin parameter $a/m$ lies in the range $0.236-0.262$, indicating moderate rotation of the BH. The resonance radius $r/m$ lies in the range $5.53- 6.22$, which is located near the inner region of the accretion disk where the oscillatory mechanisms responsible for the QPOs are expected to occur.

Furthermore, in this analysis, the magnetic field parameter $b$, which characterizes the electromagnetic contribution in the Kerr-Bertotti-Robinson spacetime, is constrained only through upper limits at the $90\%$ confidence level for all considered X-ray binary sources. The inferred peak of the dimensionless parameter $b$ is consistent with zero, and upper bounds are obtained in the range $06821 -0.09050$ at the $90\%$ confidence level. These results indicate that the magnetic field parameter $b$ slightly modifies the effective potential governing the motion of particles in the accretion disk, leading to small shifts in the epicyclic frequencies and in the resonance radius where the oscillatory modes satisfy the FR condition. The weak statistical constraint on $b$ shows that the spacetime geometry remains close to the Kerr limit, with small magnetic corrections producing only minor modifications to the particle dynamics.

Moreover, to examine the physical impact of the small values of the dimensionless magnetic field parameter $b$ within the allowed range inferred from the MCMC analysis, we analyze the behavior of test particle motion at the innermost stable circular orbit (ISCO).
For this purpose, we determine $r_{\mathrm{ISCO}}$, along with the corresponding specific energy $E_{\mathrm{ISCO}}$, specific angular momentum $L_{\mathrm{ISCO}}$, and the orbital frequency $\Omega_{\phi,\mathrm{ISCO}}$. We also analyze the energy flux and temperature profile of the accretion disk to investigate how the small values of the dimensionless magnetic field parameter influence the dynamics of particles near the considered BH.

\begin{table*}[htbp]
\begin{ruledtabular}
\caption{The best-fit values derived from the MCMC analysis for several observed QPOs in X-ray binaries within the framework of the PR and FR models are presented. The mass, spin parameter, and resonance radius $(m, a/m, r/m)$ are constrained at the $68\%$ confidence level (C.L.). The magnetic field parameter is constrained at the $68\%$ C.L. for GRO J1655-40, GRS 1915+105, H1743-322, and M82 in the case of the PR model and at the $90\%$ C.L. for the remaining X-ray binary sources. In contrast, for the FR model, the magnetic field parameter is constrained only at the $90\%$ C.L. for all considered sources.}
\label{tab:bestfit}
\begin{tabular}{ccccc}
Model & $m/(m_{\odot})$ & $a/m$ & $r/m$ & $b$ \\
\hline
\multicolumn{5}{c}{\textbf{Parametric Resonance (PR) Model}} \\
\hline
GRO J1655-40   & $4.41_{-0.12}^{+0.15}$ & $0.307^{+0.042}_{-0.052}$ & $6.96_{-0.49}^{+0.39}$ & $0.0789_{-0.011}^{+0.0086}$ \\
GRS 1915+105    & $11.63 \pm 0.39$ & $0.305 \pm 0.050$ & $6.89_{-0.46}^{+0.39}$ & $0.078\pm 0.012$ \\
XTE J1550--564  & $7.15_{-0.17}^{+0.19}$ & $0.296 \pm 0.047$ & $6.74_{0.37}^{+0.29}$ & $<0.09421$ \\
XTE J1859+226   & $9.17 \pm 0.17$ & $0.277_{-0.046}^{+0.277}$ & $6.13\pm 0.24$ & $<0.09460$ \\
H1743-322      & $8.11_{-0.24}^{+0.28}$ & $0.310 \pm 0.048$ & $6.93_{-0.48}^{+0.37}$ & $0.0804_{-0.0098}^{+0.012}$ \\
M82             & $388_{-11}^{+13}$ & $0.306 \pm 0.047$ & $6.86_{-0.48}^{+0.40}$ & $0.0763_{-0.13}^{+0.0089}$ \\
Sgr A*          & $(1.46 \pm 0.98)\times10^{6}$ & $0.295 \pm 0.049$ & $6.67_{-0.46}^{+0.39}$ & $<0.09481$ \\
\hline
\multicolumn{5}{c}{\textbf{Forced Resonance (FR) Model}} \\
\hline
GRO J1655-40   & $5.05_{-0.074}^{+0.12}$ & $0.241_{-0.045}^{+0.039}$ & $5.53 \pm 0.17$ & $<0.08924$ \\
GRS 1915+105    & $13.21 \pm 0.37$ & $0.245_{-0.044}^{+0.039}$ & $5.57 \pm 0.20$ & $<0.08323$ \\
XTE J1550-564  & $8.08_{-0.16}^{+0.21}$ & $0.246_{-0.044}^{+0.051}$ & $5.53 \pm 0.20$ & $<0.09050$ \\
XTE J1859+226   & $9.58 \pm 0.19$ & $0.255 \pm 0.043$ & $5.88^{+0.20}_{-0.23}$ & $<0.07437$ \\
H1743-322      & $9.24_{-0.21}^{+0.24} $ & $0.242_{-0.034}^{+0.044}$ & $5.53_{-0.20}^{+0.16}$ & $<0.0901$ \\
M82             & $435.5_{-10}^{+9.1}$ & $0.236 \pm 0.042$ & $5.61 \pm 0.18$ & $<0.08682$ \\
Sgr A*          & $(1.33_{-0.15}^{+0.13})\times10^{6}$ & $0.266 \pm 0.048$ & $6.22_{-0.53}^{+0.34}$ & $<0.06821$ \\
\end{tabular}
\end{ruledtabular}
\end{table*}

For this purpose, we first determine the ISCO from the effective potential governing the test particles' motion in the vicinity of the considered BH. The ISCO obtained from Eq.~(\ref{a8}) by imposing the condition
\begin{eqnarray}
\frac{d^2V_{\rm eff}}{dr^2}=0\, .
\end{eqnarray}
We are unable to determine the ISCO radius, i.e., $r_{\mathrm{ISCO}}$, analytically; therefore, it is obtained numerically. Using this value, we also compute the corresponding quantities $E_{\mathrm{ISCO}}$, $L_{\mathrm{ISCO}}$, and $\Omega_{\phi,\mathrm{ISCO}}$, which are presented in Table~\ref{tab:ISCOvalues}.

\begin{table}[ht]
\begin{ruledtabular}
\caption{We present the values of the ISCO radius and the corresponding quantities for various values of the dimensionless magnetic field parameter $b = Bm$, with $m = 1$ and $a/m = 0.2$.}
\label{tab:ISCOvalues}
\begin{tabular}{|c|c|c|c|c|}
$b$ & $r_{\mathrm{ISCO}}$ & $\tilde{E}_{\mathrm{ISCO}}$ & $\tilde{L}_{\mathrm{ISCO}}$ & $\Omega_{\phi,\mathrm{ISCO}}$ \\
\hline
0.00 & 5.3294 & 0.9353 & 3.2640 & 0.0799 \\
0.03 & 5.3344 & 0.9389 & 3.2808 & 0.0817 \\
0.06 & 5.34951 & 0.95015 & 3.3330 & 0.0868 \\
0.09 & 5.3748 & 0.9702 & 3.4257 & 0.0955 \\
\end{tabular}
\end{ruledtabular}
\end{table}

Table~\ref{tab:ISCOvalues}, shows that as the values of the dimensionless magnetic field parameter $b$ rise, the ISCO radius also increases, as well as the corresponding quantities $\Omega_{\phi,\mathrm{ISCO}}$, $E_{\mathrm{ISCO}}$, and $L_{\mathrm{ISCO}}$ increase. This behavior indicates that as the strength of the dimensionless magnetic field increases, the circular orbit at the ISCO becomes less gravitationally bound. Physically, the dimensionless magnetic field parameter $b$ modifies the effective potential governing the motion of test particles, which shifts the location of the ISCO and slightly alters the orbital dynamics. Consequently, the azimuthal frequency $\Omega_{\phi,\mathrm{ISCO}}$ is also affected by the magnetic corrections introduced in the spacetime geometry.

\section{Accretion Disk Radiation and Temperature in the Kerr-Bertotti-Robinson Spacetime}
We analyze the energy flux produced by the accretion disk around the Kerr-Bertotti-Robinson BH. The radiation emitted from the accretion disk arises due to matter accreting around the BH and is influenced by the dimensionless magnetic field parameter $b$. We study the radiation flux in the equatorial plane by using the approach provided in Refs.~\cite{Novikov1973, Page:1974he, Thorne:1974ve,Boshkayev21PRD,Alloqulov24CPC,Uktamov25PDU....4701743U}, which is given by the following expression
\begin{equation}
F(r) = -\frac{\dot{M}_0}{4\pi\sqrt{-g}\,(\tilde{E}-\Omega_{\phi}\tilde{L})^2}
\int_{r_{ISCO}}^{r} (\tilde{E}-\Omega_{\phi}\tilde{L})\,\tilde{L}_{,r}\,dr ,
\tag{2.66}
\end{equation}
where $\tilde{E}$ denotes the specific energy, $\tilde{L}$ denotes the specific angular momentum, and $\dot{M}$ is the mass accretion rate, while $\sqrt{-g}=\sqrt{-g_{rr}\left(g_{tt}g_{\phi\phi}-g_{t\phi}^{2}\right)}$ is the determinant of the metric in the equatorial plane $\theta=\pi/2$.

The matter accreting in the accretion disk is assumed to be in thermodynamic equilibrium. The energy flux emitted from the surface of the accretion disk follows the Stefan–Boltzmann law. As a result, the temperature of a geometrically thin blackbody disk can be expressed as
\begin{equation}
F(r)=\sigma T_{}^{4},
\label{flux}
\end{equation}
here, $\sigma$ represents the Stefan-Boltzmann constant, and $F(r)$ is the energy flux, while $T$ indicates the intrinsic temperature.
\begin{figure*}
\centering
\includegraphics[scale=0.5]{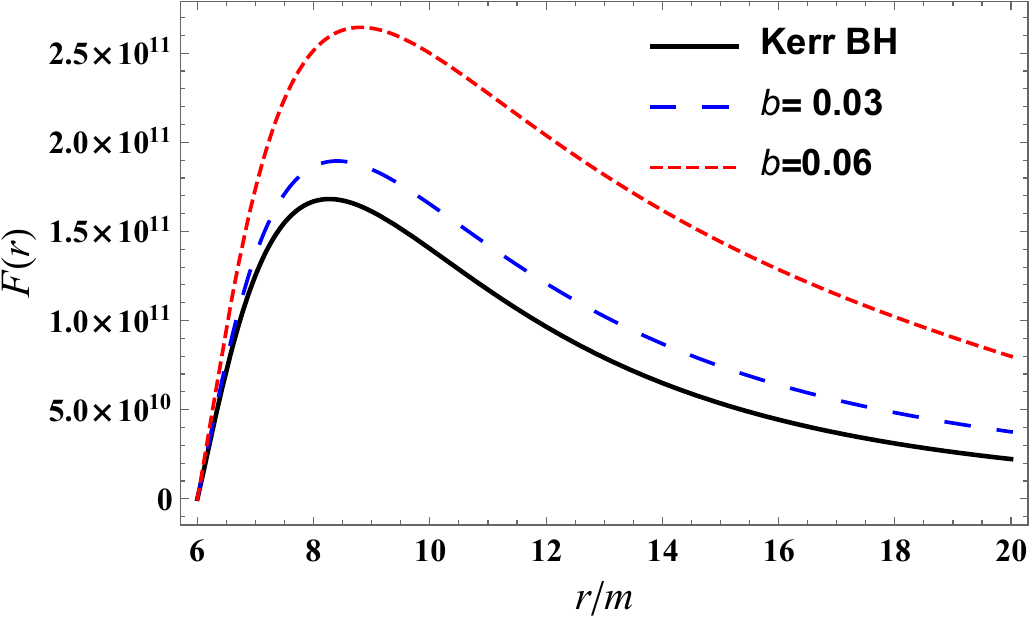}
\includegraphics[scale=0.5]{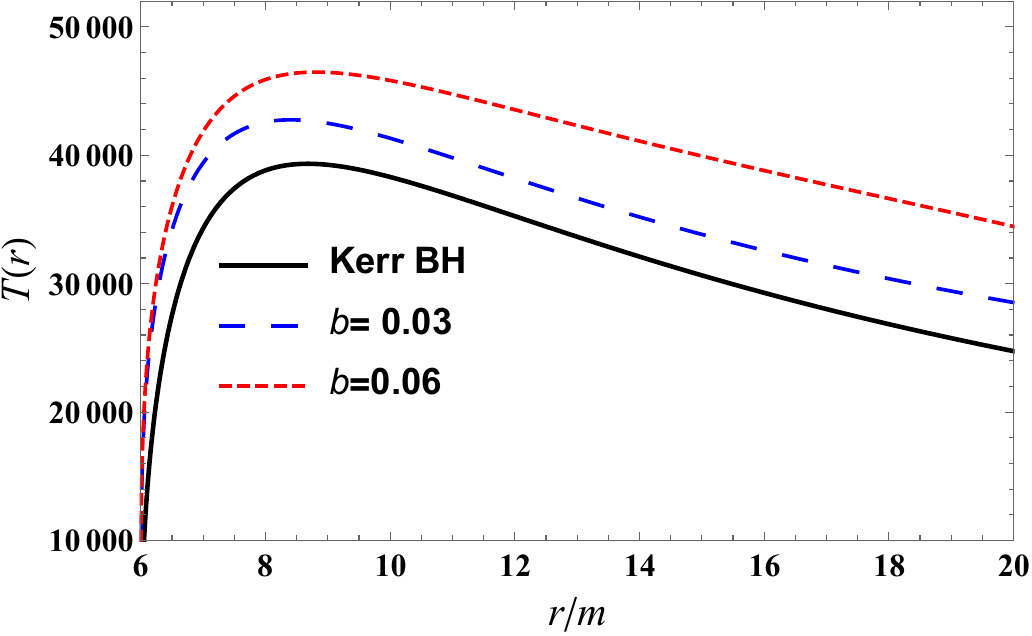}
\caption{The profiles of the energy flux $F(r)$ (left panel) and temperature $T(r)$ (right panel) of the accretion disk as functions of the dimensionless radius $r/m$ around the considered BH for mass $m=1$, spin parameter $a/m=0.1$, and different values of the dimensionless magnetic field parameter $b$.}
\label{acc}
\end{figure*}
The energy flux $F(r)$ and the temperature profile of the disk are plotted in Fig.~\ref{acc}. In the left panel of the figure, we plot the energy flux as a function of the dimensionless radial distance $r/m$. It can be seen that the maximum energy flux is observed in the central region of the accretion disk around the considered BH, and it declines as we move away from the central part of the disk. We also examine the impact of the dimensionless magnetic field parameter $b$ within the allowed range inferred from the MCMC analysis, and it is observed that as the value of the dimensionless parameter increases, the energy flux also increases. 

In the right panel of the figure, we analyze the behavior of the temperature profile of the disk. It can be seen that the temperature of the disk also increases with increasing values of $b$. The temperature attains its maximum value in the central region of the accretion disk and decreases as we move away from the central part of the disk.

\begin{figure*}
\centering
\includegraphics[scale=0.5]{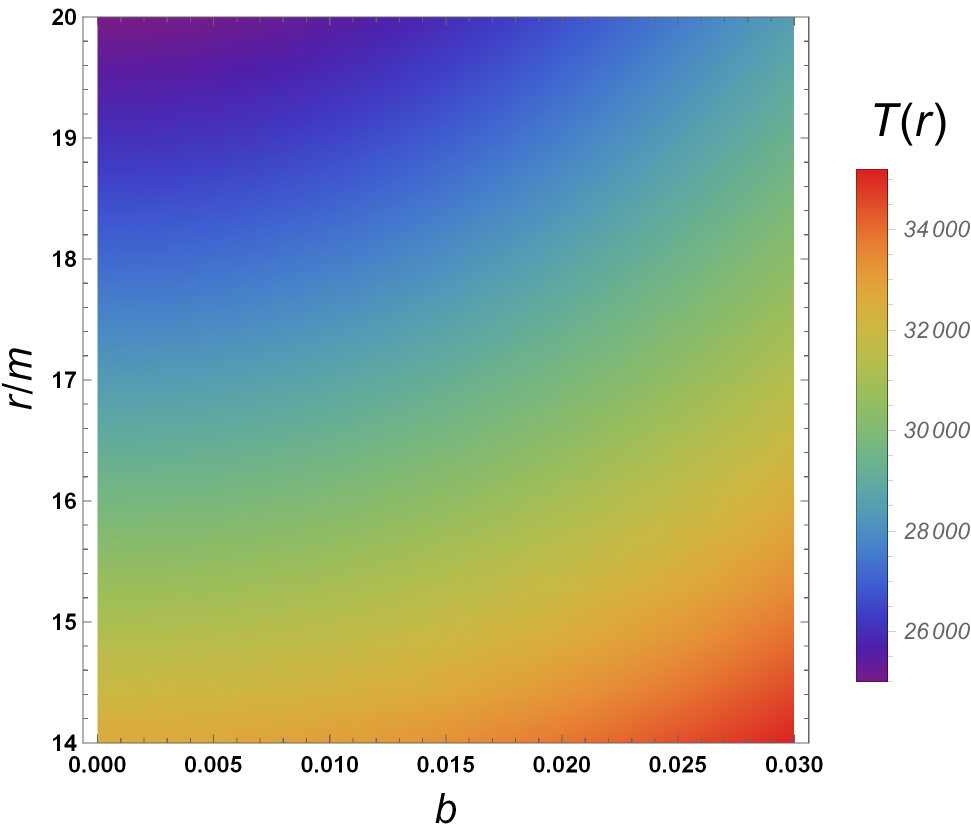}
\includegraphics[scale=0.5]{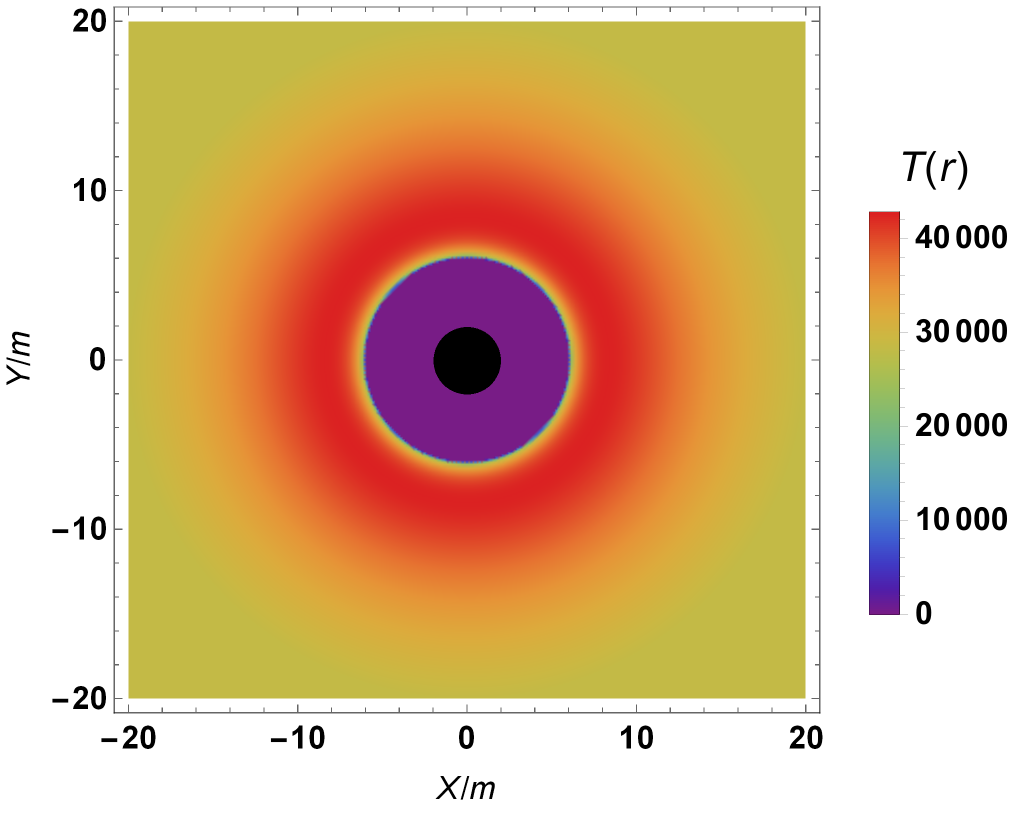}
\caption{The radial temperature profile for the dimensionless magnetic field parameter $b$ is shown in the left panel, while the corresponding density plot of the temperature distribution on the equatorial $X$--$Y$ plane is presented in the right panel. Here, $X$ and $Y$ denote the Cartesian coordinates.}
\label{Temperature}
\end{figure*}

Furthermore, for a clearer understanding, we present the color map of the temperature distribution of the accretion disk, as shown in Fig.~\ref{Temperature}. From the figure, one can observe that the dark region is located inside the inner edge of the accretion disk, whereas the red region corresponds to the maximum temperature of the disk. It is clearly seen that the maximum temperature is obtained in the inner region of the accretion disk, and it increases with increasing values of the dimensionless magnetic field parameter $b$.

\section{Conclusions}

In this study, we examine the QPOs in the spacetime of a Kerr BH immersed in a uniform Bertotti-Robinson dimensionless magnetic field. The test particle's motion is studied by deriving the geodesic equations and computing the fundamental frequencies. These frequencies are used to construct theoretical predictions for QPOs within the frameworks of the PR and FR models. By utilizing observational data obtained from QPO X-ray binaries, we constrain the parameters of the considered spacetime through MCMC simulations. 
In this analysis, we use the \texttt{GetDist} package to analyze the posterior samples, extract the marginalized constraints on the mass $m$, spin parameter $a/m$, and resonance radius $r/m$, and generate the one- and two-dimensional posterior distributions at the $68\%$ C.L. Our results show that the obtained masses are consistent with the observational data reported in Table~\ref{1a}, while the spin parameter indicates that the considered BH is moderately rotating. Moreover, the resonance radius $r/m$ is located in the inner region of the accretion disk, close to the ISCO, suggesting that the oscillatory mechanism responsible for the observed QPOs originates in the inner disk region.

An important aspect of the analysis is the constraint on the magnetic field parameter associated with the Kerr–Bertotti–Robinson spacetime. Also, for X-ray binaries GRS 1915+105, GRO J1655-40, H1743-322, and M82~X-1, we obtained non-zero values of the $b$ at $68\%$ C.L., while for the remaining sources, we obtained only an upper bound at $90\%$ C.L. for the case of the PR model. In contrast, for the case of the FR model, we only upper for the parameter $b$ at $90\%$ C.L for all consider QPO X-ray binaries listed in the Table \ref{1a}. Also, for GRO J1655-40, GRS 1915+105, H1743-322, and M82~X-1 X-ray binaries, we obtain nonzero values of the dimensionless magnetic field parameter $b$ at the $68\%$ C.L., while for the remaining sources, we obtain only upper bounds found within $0.09421-0.09481$ at the $90\%$ C.L. in the scenario of the PR model. In contrast, for the FR model, we obtain only upper bounds for $b$ within 
$0.06821- 0.09050$ at the $90\%$ C.L. for all considered QPO X-ray binaries listed in Table~\ref{1a}. The relatively weak statistical constraint on the parameter $b$ indicates that the spacetime geometry remains close to the Kerr limit, with small magnetic corrections producing only minor modifications to the particle dynamics. Our results show that the dimensionless magnetic parameter $b$ is small, but not negligible. It slightly modifies the effective potential governing the motion of particles in the accretion disk, leading to small shifts in the epicyclic frequencies and in the resonance radius where the oscillatory modes satisfy the PR and FR conditions. To analyze the physical impact of the dimensionless parameter $b$ on the motion of test particles in the accretion disk, we investigate the ISCO orbits and evaluate the specific energy, specific angular momentum, and specific angular velocity at the ISCO. We also study the energy flux and radiation temperature of the accretion disk around the considered BH within the allowed MCMC parameter range. Our results show that increasing values of the magnetic field parameter lead to a gradual increase in the ISCO radius $r_{\rm ISCO}$ as well as in the corresponding orbital quantities. Furthermore, we find that the energy flux and temperature reach their maximum values in the inner region of the disk and decrease with increasing radial distance, while even a small increase in the magnetic field slightly enhances both the energy flux and the temperature of the disk.
 
Overall, the results demonstrate that QPO observations provide a useful tool for probing the spacetime structure around BHs and for constraining the parameters associated with external electromagnetic fields. Although the magnetic contribution remains relatively small, it introduces measurable modifications to the orbital dynamics and accretion disk properties in the Kerr–Bertotti–Robinson spacetime.

\section*{Acknowledgements}

This work is supported by the National Natural Science Foundation of China under Grants No.~12275238, No.~12542053, and No.~11675143, the National Key Research and Development Program of China under Grant No. 2020YFC2201503, the Zhejiang Provincial Natural Science Foundation of China under Grants No.~LR21A050001 and No.~LY20A050002, and the Fundamental Research Funds for the Provincial Universities of Zhejiang in China under Grant No.~RF-A2019015.

\bibliography{ref1_bh}

\end{document}